\documentclass[aps,pre,reprint,superscriptaddress,nofootinbib]{revtex4-2}

\usepackage{amsmath,amssymb,bm}
\usepackage{graphicx}
\usepackage{booktabs}
\usepackage{microtype}
\usepackage{xcolor}
\usepackage{hyperref}
\hypersetup{colorlinks=true,citecolor=blue,linkcolor=blue,urlcolor=blue}
\usepackage{siunitx}

\newcommand{\E}{\mathbb{E}}

\newcommand{\I}{\mathcal{I}}
\newcommand{\dd}{\mathrm{d}}

\newcommand{\one}{\mathbf{1}}

\begin{document}

\title{Transient Supercriticality in Hawkes Processes:\\Detection, Identifiability, and Apparent Criticality}

\author{Didier Sornette}
\affiliation{Institute of Risk Analysis, Prediction and Management, Southern University of Science and Technology, Shenzhen, China}
\date{\today}

\begin{abstract}
Self-exciting (Hawkes) models of epidemics, earthquakes, financial markets, and social cascades are usually calibrated with a branching ratio $n$ below one. This is a condition for stationarity, not for existence: a process with $n>1$ grows exponentially but remains finite on any finite horizon. Imposing stationarity is also misleading, because transient supercritical episodes then masquerade as near-criticality. Reconstructing the latent genealogy of events, we show that a single finite record can reveal an excursion into the regime $n(t)>1$, and back, and date its onset and end. Detectability is governed by the excess $n-1$ times the square root of the number of exposed parents. Even when the memory time of the triggering is unknown, the data determine the growth rate of activity during the episode nearly as precisely as when it is known, and hence whether $n$ crossed unity, whereas the value of $n$ requires independent knowledge of the memory time. Letting the data choose between endogenous amplification and external forcing prevents confusing the two. In the 2015 Korean MERS outbreak, our reconstruction shows that transmission by hospital cohorts had collapsed by 28 May, before the incidence peak and two weeks before the instantaneous reproduction number fell below one; in real time, it establishes the collapse a few days earlier than this standard indicator. Our framework separates stationarity, non-explosion, and transient supercritical amplification, and provides a practical route for detecting short-lived endogenous instabilities in self-exciting systems.
\end{abstract}

\maketitle

\section{Introduction}

Many self-exciting systems appear to pass through transient phases in which each event triggers, on average, more than one further event, before the amplification subsides. Every epidemic wave begins with a reproduction number above one and ends when interventions, behavioral change, or the depletion of susceptibles bring it below one \cite{Cori2013,Flaxman2020,Garetto2021}. In seismicity, spatially resolved inversions of the epidemic-type aftershock sequence (ETAS) model for California return branching ratios $n$ up to about $1.2$ in several regions, values that can only be temporary \cite{Nandan2017}, although finite catalogues, missing events, and truncation also bias such estimates \cite{Seif2017}. In financial markets, Hawkes calibrations of high-frequency activity find an endogeneity that has risen over the years towards the critical value $n=1$ \cite{FilimonovSornette2012} or lies close to it \cite{Hardiman2013,JaissonRosenbaum2015}. On social media, a fraction of retweet cascades are estimated to be supercritical during their first minutes to hours \cite{Zhao2015}, and bursts of collective attention display the signatures of endogenous, near-critical cascades \cite{CraneSornette2008,Rizoiu2018}. In the brain, epileptic seizures share the statistics of earthquakes and have been described as departures from the near-critical dynamics of normal activity \cite{Osorio2010,Meisel2012}. More generally, transient phases of supercritical amplification are a natural mechanism for the outsized events called dragon-kings \cite{Sornette2009DK,SornetteOuillon2012}.

Yet, outside epidemiology, self-exciting models of such systems are almost always calibrated as stationary processes, with $n<1$ imposed explicitly or implied by a constant branching ratio. We argue that this is not only unnecessary on a finite horizon, as recalled below, but also misleading. A stationary fit to a record that contains a transient supercritical episode returns a branching ratio just below unity: for an episode with $n=1.35$ lasting eight memory times within a subcritical regime with $n=0.65$, the median estimate is $0.96$, against a time average of $0.79$, and criticality cannot be rejected in $86\%$ of the catalogues (Sec.~\ref{sec:detect}). A system that alternates between subcritical and supercritical phases is thus diagnosed as stationary and near-critical. This provides an alternative explanation for the branching ratios close to unity reported in many systems, and for the claims of criticality built on them \cite{Hardiman2013,Munoz2018}, in the spirit of the sweeping of an instability as an alternative to self-organized criticality \cite{Sornette1994}, and complementary to the apparent criticality produced by an unmodelled nonstationary background \cite{FilimonovSornette2015,Wheatley2019,Wehrli2021}. Allowing the data to test for transient supercriticality, rather than excluding it a priori, may therefore change conclusions drawn from stationary calibrations.

The branching ratio is the central control parameter of a linear Hawkes process: in the cluster representation, it is the expected number of direct offspring generated by one event \cite{Hawkes1971,HawkesOakes1974}.  The familiar condition $n<1$ guarantees a stationary cluster process with finite mean intensity.  It is therefore built into a large fraction of statistical theory and applied calibration, either explicitly as a parameter constraint or implicitly through a locally stationary approximation \cite{BremeaudMassoulie1996,Roueff2016,MammenMueller2023}.  It is also built into the way $n$ is commonly measured.  In a stationary process the mean intensity is $\Lambda=\mu/(1-n)$, so that the fraction of immigrants is $\mu/\Lambda=1-n$ and the fraction of triggered events is exactly $n$.  The branching ratio is therefore often estimated and interpreted as the fraction of endogenous events, $n=1-\mu/\Lambda$, for instance as a measure of the endogeneity, or reflexivity, of financial markets \cite{FilimonovSornette2012}, or as the fraction of events attributed to triggering by stochastic declustering \cite{ZhuangOgataVJ2002}.  Such an estimate is bounded by unity by construction, since a fraction cannot exceed one.  The identification of the two quantities, however, holds only in the stationary regime.  In a finite window that contains an interval with $n(t)>1$, the fraction of triggered events remains below one while the mean number of offspring per parent exceeds one: in the catalogues of Sec.~\ref{sec:detect} with an episode $n_+=1.35$, triggered events make up a median $92\%$ of all events, a value that would be read as a near-critical branching ratio.  As a result, a mathematical condition for \emph{stationarity} has become an inferential prior, enforced either by a parameter constraint or by the very definition of the estimated quantity: the model is prevented from asking whether the data contain a finite interval with $n(t)>1$.  We instead estimate $n$ as the mean number of offspring per exposed parent, a quantity that is not bounded by unity.

The restriction of stationarity is stronger than finite-time well-posedness requires. In point-process terminology, ``explosion'' means the accumulation of infinitely many events before a finite time. For a standard linear Hawkes process with a regular exponential kernel and finite mean fertility, \(n>1\) destroys stationarity but does not by itself imply finite-time explosion; instead, the mean activity grows asymptotically at a positive Malthusian rate. Nonstationary and time-varying Hawkes processes have been studied theoretically and statistically \cite{TannenbaumBurak2017,Deschatre2025}, although their inference is generally developed under stability or local-stationarity assumptions, while supercritical Hawkes dynamics has also become an explicit object of mathematical analysis \cite{LiuXuZhang2026}. In the ETAS context, Helmstetter and Sornette showed that a finite branching ratio \(n>1\) produces an exponentially accelerating seismicity rate at sufficiently long times \cite{HelmstetterSornette2002}. An even more singular regime arises when event fertilities are power-law distributed with tail index below unity, so that the mean fertility and formally the branching ratio, diverges. Sornette and Helmstetter showed that the resulting anomalous branching dynamics can generate genuine stochastic finite-time singularities of the activity rate \cite{SornetteHelmstetter2002PRL}. These results show that the conventional boundary \(n=1\) separates stationary from supercritical branching dynamics, rather than admissible from inadmissible finite-time realizations. This motivates the inference problem considered here: to determine from a single finite realization whether the process enters a transient regime with \(n(t)>1\), and to estimate its onset, duration, and amplitude while discriminating endogenous amplification from exogenous forcing.

There is substantial work on nonstationary Hawkes processes and time-varying reproduction functions \cite{Roueff2016,MammenMueller2023,Deschatre2025,TannenbaumBurak2017}.  Much of the accompanying asymptotic theory nevertheless relies on a local or uniform subcriticality condition to control moments and obtain stationary approximations.  The transition through $n=1$ is precisely where that approximation ceases to be appropriate.  We therefore ask a narrower and complementary question: given an observation window $[0,T]$, can one infer a finite episode
\begin{equation}
 n_1<1\quad\longrightarrow\quad n_+>1\quad\longrightarrow\quad n_3<1,
 \label{eq:target}
\end{equation}
and estimate its onset, duration, and amplitude while separating endogenous reproduction from exogenous immigration?

We formulate the inference of such episodes as a latent-genealogy problem. Section~\ref{sec:finite} recalls why $n>1$ violates stationarity but not finite-horizon existence, derives the Malthusian relation $n=1+\gamma/\beta$ between the branching ratio and the exponential growth rate $\gamma$ of the event population for an exponential kernel of memory time $1/\beta$, and shows that local growth-rate estimates from a single record are too noisy to locate the crossing. Section~\ref{sec:em} develops the estimator. Posterior probabilities that each event is an immigrant or the child of a specific predecessor \cite{ZhuangOgataVJ2004,VeenSchoenberg2008} yield closed-form expectation--maximization updates of piecewise-constant branching ratios indexed by parent time, with the exact correction for descendants that fall beyond the end of the observation window; the onset and termination of the episode are selected by an information criterion, the memory time is estimated by profile likelihood, $n_+>1$ is tested by a likelihood ratio, and the transient growth rate serves as an independent cross-check.

Section~\ref{sec:results} shows on simulated catalogues that a supercritical episode is recovered without any finite-time explosion, and establishes the detectability phase diagram: power depends jointly on the excess $n_+-1$ and on the duration of the episode in memory times, and both enter through a single variable, the excess multiplied by the square root of the offspring exposure of the episode. When the memory time is estimated, a short record identifies the growth exponent $\beta(n_+-1)$ rather than $n_+$ itself, so that the sign of $n_+-1$ remains testable whereas its magnitude requires independent information on the kernel. Section~\ref{sec:endoexo} addresses the main confounder, a surge of exogenous immigration. Letting the likelihood choose between an endogenous episode in $n(t)$ and an exogenous episode in the background rate keeps false detections under a pure exogenous surge below $1\%$ at a modest cost in power, whereas kernel-smoothed stochastic reconstructions of the background and of the branching ratio \cite{ZhuangOgataVJ2002,ZhuangMateu2019} introduce a smoothing scale that competes with the change points and, depending on the bandwidth, either absorb genuine supercritical bursts into the background or report spurious crossings of $n=1$. The same section shows that stationary catalogues close to or at criticality do not produce false detections, and that a misspecified memory time can turn an exogenous surge into a spurious episode. Section~\ref{sec:mers} applies the method to the 2015 Korean MERS outbreak. A constant-$n$ fit suggests sustained supercriticality, whereas the time-resolved reconstruction separates the index superspreader, a supercritical cohort of hospital transmission ($n\simeq3.9$), and subcritical parent cohorts from 28 May onward, in agreement with contact-tracing data and with standard case reproduction numbers. The incidence peak is the delayed echo of the supercritical cohort, and the instantaneous reproduction number falls below unity about two weeks after the crossing; in real time, the crossing becomes testable a few days before the instantaneous estimate does. Section~\ref{sec:discussion} discusses the conditions under which transient supercriticality is identifiable from a finite record (sufficient information, separation between the memory time and the scale of background variations, and control of the kernel), and Sec.~\ref{sec:conclusion} concludes. The Appendix gives the growth rate for an Omori kernel.

\section{Stationarity versus finite-time existence}\label{sec:finite}

\subsection{Growth Exponent $\gamma$ and Branching Ratio $n=1+\gamma/\beta$}

\begin{figure}[t]
\includegraphics[width=\columnwidth]{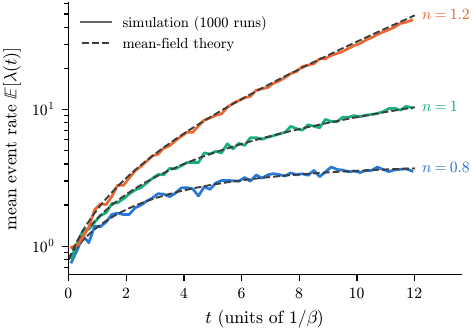}
\caption{\label{fig:finite}Stationarity versus finite-time existence.  Mean event rate from 1000 branching-process realizations for $\mu=0.8$, $\beta=1$, and $n=0.8,1,1.2$ (solid curves) compared with Eq.~\eqref{eq:meanrate} (dashed).  The supercritical case grows exponentially but is finite throughout the finite window.}
\end{figure}

Consider the univariate Hawkes intensity
\begin{equation}
 \lambda(t)=\mu+\sum_{t_i<t} n(t_i)h_\beta(t-t_i),
~
 h_\beta(u)=\beta e^{-\beta u}\one_{u>0},
 \label{eq:model}
\end{equation}
where $\mu>0$ is the immigrant rate and $n(t_i)$ is the mean number of direct offspring generated by a parent occurring at $t_i$.  The kernel is normalized, $\int_0^\infty h_\beta(u)\dd u=1$, so $n$ is directly the branching ratio.

For constant $n$, the cluster representation is a Galton--Watson genealogy embedded in continuous time.  A single immigrant has an expected $n^k$ descendants in generation $k$.  For the exponential kernel, the birth time of a $k$th-generation descendant along a lineage is Erlang distributed \cite{Erlang1909}, and
\begin{equation}
 \int_0^T h_\beta^{*k}(t)\dd t
 \le \frac{(\beta T)^k}{k!}.
\end{equation}
Hence the expected number of descendants born before any finite $T$ is bounded by
\begin{equation}
 \sum_{k\ge0} n^k\int_0^T h_\beta^{*k}(t)\dd t
 \le e^{n\beta T}<\infty .
 \label{eq:finitebound}
\end{equation}
With a finite immigrant rate, the number $N(T)$ of events in $[0,T]$ has expectation
$\E N(T)\le\mu T\,e^{n\beta T}<\infty$, which excludes finite-time explosion.
The same comparison applies to a bounded time-dependent $n(t)\le n_{\max}<\infty$.  Thus $n>1$ is compatible with a perfectly finite realization on every bounded window.

\begin{figure*}[t]
\includegraphics[width=0.94\textwidth]{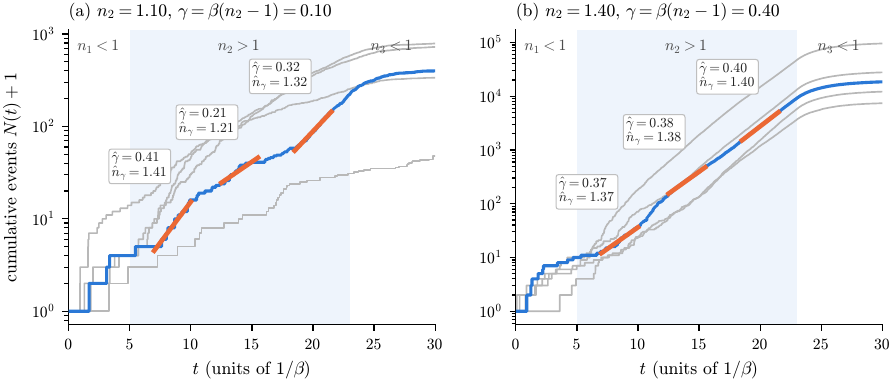}
\caption{
Local growth-rate estimates in transiently supercritical Hawkes processes.
Five independent realizations are shown in each panel as the cumulative
number of events $N(t)+1$ on a logarithmic vertical scale. The branching
ratio follows the three-regime profile of Eq.~\eqref{eq:target}, with
$n_1=n_3=0.65$, $t_1=5$, $t_2=23$, and $T=30$, so that the
supercritical regime occupies $60\%$ of the observation interval.
The triggering kernel is exponential,
$g(t)=\beta e^{-\beta t}$, with $\beta=1$, and the background
rate is $\mu=0.7$. Panel (a) has $n_2=1.10$, corresponding to the
asymptotic Malthusian exponent $\gamma=\beta(n_2-1)=0.10$, while
panel (b) has $n_2=1.40$, corresponding to $\gamma=0.40$.
For the realization with the median final size in each panel (highlighted), the three short
straight segments are local linear fits to $\log[N(t)+1]$ over windows of three memory times at three
different times within the supercritical regime. Each segment is
labelled by its estimated slope $\hat\gamma$ and the corresponding
branching-ratio estimate
$\hat n_\gamma=1+\hat\gamma/\beta$.
For $n_2=1.10$, the three local estimates are
$\hat n_\gamma=1.41$, $1.21$, and $1.32$, despite the true value
$n_2=1.10$. For $n_2=1.40$, they are $1.37$, $1.38$, and $1.40$,
approaching the true value only as the supercritical growth mode becomes
dominant.
}
\label{fig:local_gamma}
\end{figure*}

What changes at $n=1$ is the \emph{long-time} behavior of the mean activity; whether the process itself admits a stationary version at criticality is a subtler question, whose answer depends on the tails of the memory kernel and of the fertility distribution \cite{Sornette2026CriticalRF}.  Let $m(t)=\E[\lambda(t)]$ for a process started empty at $t=0$.
Taking expectations in Eq.~\eqref{eq:model} gives
\begin{equation}
 m(t)=\mu+n\beta\int_0^t e^{-\beta(t-s)}m(s)\dd s,
\end{equation}
which is equivalent to
\begin{equation}
 \dot m=\beta(n-1)m+\beta\mu,
 \qquad m(0)=\mu.
 \label{eq:meanode}
\end{equation}
Therefore
\begin{equation}
 m(t)=
 \begin{cases}
 \displaystyle
 \mu\,\frac{n e^{\beta(n-1)t}-1}{n-1}, & n\ne1,\\[6pt]
 \mu(1+\beta t), & n=1.
 \end{cases}
 \label{eq:meanrate}
\end{equation}
For $n<1$, $m(t)$ converges to the stationary mean $\mu/(1-n)$; for $n=1$ it grows linearly; for $n>1$ it grows exponentially but remains finite at every finite time.  
Figure~\ref{fig:finite} illustrates Eq.~\eqref{eq:meanrate} by direct branching simulations.

For a general normalized triggering kernel $h(u)$, a supercritical mean-field mode $m(t)\propto e^{\gamma t}$ satisfies the Malthusian equation
\cite{HawkesOakes1974,Nerman1981}.
\begin{equation}
 1=n\int_0^\infty e^{-\gamma u}h(u)\dd u.
 \label{eq:malthus}
\end{equation}
For the exponential kernel, Eq.~\eqref{eq:malthus} reduces to
\begin{equation}
 \gamma=\beta(n-1),\qquad
 n=1+\frac{\gamma}{\beta}.
 \label{eq:growth}
\end{equation}
Equation~\eqref{eq:growth} provides a second, genealogy-free estimator of $n$ when a sufficiently long supercritical growth segment is visible. 
Estimating \(\gamma\) from a single finite realization is intrinsically noisy. Early-time dynamics are dominated by branching fluctuations and preasymptotic effects; near \(n=1\), the small \(\gamma\) further delays the emergence of a measurable growth mode. At the opposite extreme, strong supercriticality rapidly generates very large activity, so that finite-size effects, nonlinear saturation, observational constraints, or the finite duration of the episode distort the ideal exponential regime.  
These limitations motivate the need for genealogical inference of the branching structure, rather than relying on growth-rate estimation alone.

\subsection{Local Growth-Rate Estimates and Their Variability}

The difficulty of using the Malthusian growth rate as an estimator is
illustrated in Fig.~\ref{fig:local_gamma}. Even when the supercritical
regime occupies more than half of the observation window, the apparent
slope of $\log[N(t)+1]$ is strongly time dependent in a single finite
realization. For the weakly supercritical case $n_2=1.10$, whose
asymptotic value is only $\gamma=0.10$, three local fits within the same
supercritical episode yield $\hat\gamma=0.41$, $0.21$, and $0.32$,
corresponding to inferred branching ratios $1.41$, $1.21$, and $1.32$.
Thus stochastic branching fluctuations and preasymptotic dynamics can
produce errors comparable to, or larger than, the distance from
criticality itself.

For the more strongly supercritical case $n_2=1.40$, the asymptotic
growth mode is more clearly expressed: successive local estimates give
$\hat n_\gamma=1.37$, $1.38$, and $1.40$, progressively approaching the
true value. However, this improved asymptotic behavior comes at the price
of a very rapid increase of the event population and a correspondingly
large dispersion between realizations. Hence there is an intrinsic
tension in growth-rate inference: weak supercriticality is obscured by
stochastic and preasymptotic fluctuations, whereas strong
supercriticality rapidly generates extremely large activity and becomes
sensitive to finite-size, finite-window, and observational constraints.
This motivates the genealogical inference developed below, which
estimates the branching structure directly rather than inferring it
solely from the macroscopic growth rate.

\section{Genealogical expectation-maximization (EM) inference}\label{sec:em}

\subsection{Latent parent probabilities}

Let $0<t_1<\cdots<t_N<T$ be observed event times.  We approximate $n(t)$ by $R$ plateaus $n_r$ over parent-time intervals $I_r$.  Define
\begin{equation}
 H_{jr}=\sum_{i<j:\,t_i\in I_r}h_\beta(t_j-t_i),
 \qquad
 \lambda_j=\mu+\sum_{r=1}^{R}n_rH_{jr}.
 \label{eq:design}
\end{equation}
The posterior probability that event $j$ is an immigrant is
\begin{equation}
 p_{0j}=\frac{\mu}{\lambda_j},
 \label{eq:p0}
\end{equation}
while the probability that $j$ is a direct child of $i<j$ is
\begin{equation}
 p_{ij}=\frac{n_{r(i)}h_\beta(t_j-t_i)}{\lambda_j}.
 \label{eq:pij}
\end{equation}
These are the standard stochastic-genealogy probabilities \cite{ZhuangOgataVJ2004,VeenSchoenberg2008}, which organize the incomplete-data likelihood around the Hawkes branching structure.

A parent at $t_i$ is observed only through descendants born before $T$.  Its observable offspring fraction is
\begin{equation}
 F_i=\int_0^{T-t_i}h_\beta(u)\dd u
 =1-e^{-\beta(T-t_i)}.
 \label{eq:Fi}
\end{equation}
For regime $r$, let us define the posterior offspring count and exposure
\begin{align}
 Q_r&=\sum_{t_i\in I_r}\sum_{j>i}p_{ij}
 =\sum_j\frac{n_rH_{jr}}{\lambda_j},\\
 E_r&=\sum_{t_i\in I_r}F_i.
 \label{eq:exposure}
\end{align}
The M step of the genealogical expectation-maximization (EM) estimator is then
\begin{equation}
 \mu^{\rm new}=\frac{1}{T}\sum_jp_{0j},
 \qquad
 n_r^{\rm new}=\frac{Q_r}{E_r}.
 \label{eq:EM}
\end{equation}
For fixed $\beta$ and fixed regime boundaries, Eq.~\eqref{eq:EM} is the exact EM update for the piecewise-constant model.  In particular, the edge correction acts on the aggregate exposure $E_r$ rather than multiplying individual offspring counts by $1/F_i$, which would have large variance near $T$.

\paragraph*{Relation to reproduction-number estimators.}  In epidemiology, the same attribution underlies the estimation of reproduction numbers from symptom-onset dates.  Without immigration and with a constant $n$, Eq.~\eqref{eq:pij} reduces to $p_{ij}=h(t_j-t_i)/\sum_{k<j}h(t_j-t_k)$, the relative likelihood that case $i$ infected case $j$, and $\sum_{j>i}p_{ij}$ is the case reproduction number of Wallinga and Teunis \cite{WallingaTeunis2004}; the exposure $F_i$ plays the role of the right-truncation correction introduced for real-time use by Cauchemez \emph{et al.}~\cite{Cauchemez2006}.  The EM update \eqref{eq:EM} iterates this attribution with regime-specific fertilities, which turns it into a maximum-likelihood estimator of piecewise-constant case reproduction numbers, and it adds immigration, change-point selection, and an explicit test of $n>1$.  The branching ratio $n(t_i)$ is indexed by the time of the parent: it is a case, or cohort, reproduction number.  Renewal-equation estimators instead target the instantaneous reproduction number $R(t)$, indexed by the time of the offspring, i.e.\ $\lambda(t)=R(t)\sum_{t_i<t}h(t-t_i)$ in Hawkes form \cite{Cori2013}; the two are related by a convolution with the generation-interval distribution and are offset by about one mean generation interval \cite{Fraser2007}.  Section~\ref{sec:mers} compares the three estimators on the same outbreak.

The observed log likelihood is
\begin{equation}
 \ell(\mu,\bm n)=
 \sum_{j=1}^N\log\lambda_j-\mu T-\sum_{r=1}^{R}n_rE_r.
 \label{eq:ll}
\end{equation}
For a fixed segmentation, the negative Hessian is
\begin{equation}
 \I=\sum_{j=1}^{N}\frac{\bm x_j\bm x_j^{\mathsf T}}{\lambda_j^2},
 \qquad
 \bm x_j=(1,H_{j1},\ldots,H_{jR})^{\mathsf T},
 \label{eq:info}
\end{equation}
which we use for local standard errors.  Conditional on a known candidate interval, a one-sided diagnostic for supercriticality is
\begin{equation}
 Z=\frac{\widehat n_+-1}{\operatorname{se}(\widehat n_+)}.
 \label{eq:wald}
\end{equation}
Here $n_+$ denotes the plateau $n_r$ of the candidate supercritical regime $I_r=[a,b)$, i.e., the middle plateau of the three-regime alternative \eqref{eq:target}. Thus $\widehat n_+$ is its estimate at the EM fixed point \eqref{eq:EM}, which is the maximum-likelihood estimate, and $\operatorname{se}(\widehat n_+)=[(\I^{-1})_{r+1,\,r+1}]^{1/2}$ is its standard error, obtained from the diagonal element of the inverse observed information \eqref{eq:info} evaluated at that fixed point; the index is shifted by one because the first component of $\bm x_j$ refers to $\mu$.
We use $Z>1.645$ as a nominal one-sided 5\% detection threshold.  This fixed-window test does not include change-point search uncertainty; unknown intervals are treated separately below by model selection.

\subsection{Unknown supercritical interval}
\label{sec:interval}

For the three-regime alternative in Eq.~\eqref{eq:target}, we scan candidate onset and termination times $(a,b)$ on a grid and run Eq.~\eqref{eq:EM} to convergence for each pair.  The best interval minimizes
\begin{equation}
 \mathrm{BIC}(a,b)=-2\ell(\widehat\theta_{a,b})+k\log N,
 \label{eq:bic}
\end{equation}
where $\theta=(\mu,n_1,n_+,n_3)$ collects the parameters of the three-regime model, with $\beta$ added when it is estimated (Sec.~\ref{sec:beta}), the parent-time regimes being $I_1=[0,a)$, $I_2=[a,b)$, and $I_3=[b,T)$; $\widehat\theta_{a,b}$ is the maximum-likelihood estimate of $\theta$ obtained by iterating Eq.~\eqref{eq:EM} to convergence with the change points held at $a$ and $b$; $\ell$ is the log likelihood \eqref{eq:ll}; and $N$ is the number of events.  The penalty counts $k=6$ parameters: $\mu$, the three plateau values, and the two change points. The constant-$n$ null has $k=2$; when $\beta$ is estimated (Sec.~\ref{sec:beta}) both counts increase by one.  

More scalable dynamic-programming segmentation \cite{Killick2012} or fused-lasso variants can replace the grid scan when many changes are allowed, but the two-change formulation isolates the scientific question of a finite excursion through $n=1$.

\subsection{Unknown memory time and likelihood-ratio test}
\label{sec:beta}

The memory time $1/\beta$ is rarely known in applications, and a misspecified $\beta$ distorts the attribution of offspring (Sec.~\ref{sec:robust}).  We therefore include $\beta$ in the calibration by profile likelihood,
\begin{equation}
 \widehat\beta=\arg\max_\beta\,\ell\bigl(\widehat\mu(\beta),\widehat{\bm n}(\beta),\beta\bigr),
 \label{eq:profile}
\end{equation}
where $(\widehat\mu(\beta),\widehat{\bm n}(\beta))$ is the fixed point of the EM update \eqref{eq:EM} at given $\beta$, obtained by a bounded one-dimensional search over $\log\beta$; at each trial value, the EM update is iterated to convergence, and since the log likelihood \eqref{eq:ll} is concave in $(\mu,\bm n)$ for fixed $\beta$ and fixed change points, its fixed point does not depend on the initialization. The search is restricted to $\beta\in[0.2,5]$ in units of the simulated value; the lower bound is reached in $5.5\%$ of the catalogues of Sec.~\ref{sec:detect}, a manifestation of the ridge discussed below. For an unknown window, the scan of Sec.~\ref{sec:interval} and the profile \eqref{eq:profile} are alternated until $(\widehat a,\widehat b)$ is stable, which takes one or two passes.

The data constrain the growth exponent $\gamma_+=\beta(n_+-1)$ of the episode much better than $\beta$ and $n_+$ separately: the likelihood has a ridge along which $\widehat\beta$ and $\widehat n_+-1$ vary inversely [Fig.~\ref{fig:transient}(c)].  A short transient record is dominated by its macroscopic rates, the growth rate $\beta(n_+-1)$ inside the episode and the relaxation rates $\beta(1-n_r)$ outside it, whereas $\beta$ alone is informed only by the lag structure of the clustering, which is diluted in a burst where each event has many candidate parents.  The magnitude of $\widehat n_+$ therefore inherits the uncertainty on $\beta$, but its position relative to unity does not, since $n_+>1$ if and only if $\gamma_+>0$ for any $\beta>0$.  A Wald statistic such as Eq.~\eqref{eq:wald}, computed in the $n_+$ parametrization, is not invariant along this ridge and loses power.  We therefore test $H_0:\,n_+\le1$ with the signed root of the likelihood ratio,
\begin{equation}
 r=\operatorname{sgn}(\widehat n_+-1)\sqrt{2\bigl[\ell(\widehat\theta)-\ell(\widehat\theta_0)\bigr]},
 \label{eq:lr}
\end{equation}
where $\widehat\theta_0$ maximizes the likelihood over all other parameters, $\beta$ included, with $n_+$ held at unity in the M step.  Under $H_0$, $r$ is asymptotically standard normal; in short catalogues it is overdispersed, and its threshold is best calibrated by a parametric bootstrap (Sec.~\ref{sec:detect}).  We declare supercriticality when $r>1.645$ and, for an unknown window, when the transient model is also preferred by BIC.  In Sec.~\ref{sec:results} we first treat $\beta$ as known, which isolates the difficulty of locating and testing the episode, and then quantify the effect of estimating it (Secs.~\ref{sec:recover}, \ref{sec:detect}, and~\ref{sec:robust}).

\subsection{Independent growth-rate diagnostic}

Within a candidate supercritical interval, we also fit the binned rate to
\begin{equation}
 r(t)\simeq C e^{\gamma(t-a)}+D,
 \label{eq:growthfit}
\end{equation}
then use Eq.~\eqref{eq:growth}.  For the exponential kernel, this form is exact for the conditional mean rate.  Because the intensity \eqref{eq:model} is Markovian for an exponential kernel, the argument leading to Eq.~\eqref{eq:meanode} applies inside a constant-$n_+$ window irrespective of the history before $a$: conditional on the information $\mathcal F_a$ available at the onset, $\dot m=\beta(n_+-1)m+\beta\mu$, hence
\begin{equation}
 \E[\lambda(t)\,|\,\mathcal F_a]=\bigl[\lambda(a)-D\bigr]e^{\gamma(t-a)}+D,
 ~
 D=\frac{\mu}{1-n_+},
 \label{eq:growthexact}
\end{equation}
for $a\le t<b$.  The influence of pre-episode parents does not appear as a separate decaying term; it is entirely contained in $\lambda(a)$.  The offset $D$ is \emph{negative} in the supercritical regime and must therefore be left unconstrained in sign: treating it as a non-negative background offset biases $\widehat\gamma$ upward (Sec.~\ref{sec:detect}).  The fitted pair also provides an internal consistency check, $\widehat D\simeq-\beta\widehat\mu/\widehat\gamma$.  This estimator does not use the latent genealogy.  Agreement between the two methods is therefore informative; disagreement is also diagnostic because exogenous growth, a short interval, or kernel misspecification can contaminate Eq.~\eqref{eq:growthfit}.

\section{Results}
\label{sec:results}

\subsection{Recovering a supercritical episode}
\label{sec:recover}

\begin{figure*}[t]
\includegraphics[width=0.94\textwidth]{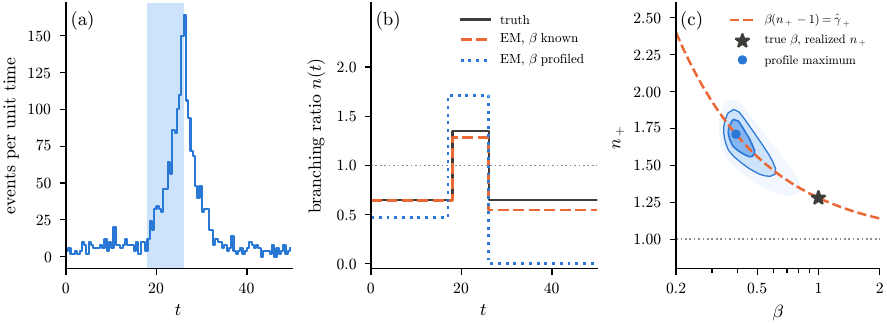}
\caption{\label{fig:transient} Recovery of a transient supercritical episode from a single finite
Hawkes realization. The branching ratio is
$n(t)=0.65$ for $t<18$, $n(t)=1.35$ for $18\leq t<26$, and
$n(t)=0.65$ for $t\geq26$, with $\mu=2$, $\beta=1$, and $T=50$; the catalogue shown ($N=1072$) is the one closest to the median size among 400 realizations.
(a) Event rate in bins of width $0.5$; the shaded region indicates
the true supercritical interval $[18,26)$.
(b) True branching-ratio profile (solid line), genealogical EM estimate
with $\beta=1$ known and the true change points supplied (dashed line), giving
$(\hat n_1,\hat n_+,\hat n_3)=(0.642,1.285,0.546)$, and EM estimate with $\beta$ profiled and the window found by a blind scan (dotted line), giving $[\hat a,\hat b)=[17,26)$, $\hat\beta=0.39$, and $(\hat n_1,\hat n_+,\hat n_3)=(0.47,1.71,0.00)$.  With $\beta$ known, the blind scan selects the same window and gives $\hat n_+=1.279$.  The horizontal dotted line marks the critical threshold $n=1$.
(c) Profile likelihood of $(\beta,n_+)$ on the selected window, all other parameters maximized: the two contour lines delineate regions $2\Delta\ell<2.30$ and $<5.99$ (approximate 68\% and 95\% joint levels); maximum (dot), and hyperbola of constant growth exponent $\beta(n_+-1)=\hat\gamma_+=0.280$ (dashed).  The star marks the true $\beta=1$ with the offspring production $n_+=1.279$ realized in this catalogue; it lies on the ridge.}
\end{figure*}

We simulate Eq.~\eqref{eq:model} exactly from its branching representation, with $\mu=2$, $\beta=1$, $T=50$, and
\begin{equation}
 n(t)=
 \begin{cases}
 1.35,&18\le t<26,\\
 0.65,&\text{otherwise}.
 \end{cases}
 \label{eq:repscenario}
\end{equation}
Catalogue sizes are broadly distributed (median $1074$ and 5--95\% range $471$--$2187$ events over 400 realizations); the displayed realization is the one closest to the median, with $N=1072$ events, and it develops a large but finite burst [Fig.~\ref{fig:transient}(a)].  If the true interval is supplied, genealogical EM gives
\begin{equation}
 (\widehat n_1,\widehat n_+,\widehat n_3)
 =(0.642,1.285,0.546),
 \label{eq:repfit}
\end{equation}
with $Z=4.89$ [Eq.~\eqref{eq:wald}] for $n_+>1$.  Using the actual simulated genealogy as an oracle, i.e.\ dividing the number of observed children of the parents of each regime by their exposure $E_r$, gives $(0.722,1.279,0.605)$.  The offspring production realized inside the episode is therefore $1.279$, below the nominal $1.35$, and the EM estimate tracks this realized value rather than the nominal parameter: the deviation from $1.35$ is finite-realization noise, not a bias of the latent allocation.  In the two subcritical regimes, the EM attributes slightly more events to immigration than the true genealogy does ($\widehat\mu=2.72$ against $\mu=2$), which lowers $\widehat n_1$ and $\widehat n_3$; this is the usual immigrant--offspring ambiguity at low activity and leaves the supercritical plateau unaffected.

When the interval is unknown, a blind integer-grid scan over onset and termination times selects $(\widehat a,\widehat b)=(17,26)$ and returns $\widehat n_+=1.279$ with $Z=4.84$ [Fig.~\ref{fig:transient}(b)].  The BIC improvement relative to the best constant-$n$ fit is
\begin{equation}
 \Delta\mathrm{BIC}=\mathrm{BIC}_{\rm transient}-\mathrm{BIC}_{\rm constant}
 =-59.8 .
\end{equation}
The constant-$n$ fit itself gives $\widehat n=0.95$: averaging over the whole window would classify this catalogue as subcritical.  The independent growth-rate fit of Eq.~\eqref{eq:growthfit} on the selected window gives $1+\widehat\gamma/\beta=1.32$: it correctly identifies supercritical growth, but, as shown below, it is markedly less precise than the genealogy-based estimate across realizations.

With $\beta$ profiled (Sec.~\ref{sec:beta}), the blind scan selects the same window $[17,26)$, with $\Delta\mathrm{BIC}=-51.0$ and $r=6.4$ [Eq.~\eqref{eq:lr}], but at $\widehat\beta=0.39$.  The fitted memory is 2.5 times too long, and the branching ratios move along the ridge to $(\widehat n_1,\widehat n_+,\widehat n_3)=(0.47,1.71,0.00)$ with $\widehat\mu=4.3$: the relaxation after the episode, produced with $\beta=1$ by subcritical cascades with $n_3=0.65$, is now carried by the slowly decaying offspring tail of the episode itself [Fig.~\ref{fig:transient}(b)].  The growth exponent is nevertheless preserved, $\widehat\beta(\widehat n_+-1)=0.280$ against the realized $1.279-1=0.279$.  The joint likelihood region [Fig.~\ref{fig:transient}(c)] is elongated along the hyperbola $\beta(n_+-1)=\widehat\gamma_+$; at the 95\% level it spans $\beta\in[0.36,0.61]$ and $n_+\in[1.44,1.84]$, entirely above $n_+=1$, but it excludes the true memory time: at this sample size, asymptotic likelihood regions for $\beta$ are not reliable (Sec.~\ref{sec:robust}).  The constant-$n$ model with $\beta$ profiled gives $\widehat n=0.94$ at $\widehat\beta=1.84$, again subcritical.  In this catalogue, the episode and its sign are therefore recovered without knowledge of the kernel, but not the magnitude of $n_+$.

To check that this recovery is typical, we repeated the blind scan on 200 catalogues with $\mu=1.5$, $\beta=1$, $T=40$, $n_0=0.65$, and $n_+=1.35$ on $[15,23)$, scanning a grid of step 2 refined to step $0.5$ around the optimum, without supplying the true window.  A transient model had lower BIC than the constant-$n$ model in $86.5\%$ of the catalogues and $\Delta\mathrm{BIC}<-10$ in $76\%$ (median $\Delta\mathrm{BIC}=-36.4$).  The termination was located exactly in $56\%$ of the catalogues and within one time unit in $87\%$; the onset was within one time unit in $54\%$.  The median $\widehat n_+$ was $1.30$ (interquartile range $1.20$--$1.37$), and a transient model with $Z>1.645$ was selected in $78\%$ of the catalogues.  The onset is systematically harder to locate than the termination because the first memory times of a supercritical episode produce few events, whereas the termination coincides with the peak of activity.  With the likelihood-ratio test \eqref{eq:lr} in place of $Z$, the detection rate is the same ($79\%$ on 100 independent catalogues); when $\beta$ is also profiled, it falls to $65\%$, which is the price of not knowing the kernel (Sec.~\ref{sec:robust}).

\subsection{Detectability is controlled by height and duration}
\label{sec:detect}

We next quantify the finite-sample detection boundary.  We fix $\mu=1.5$, $\beta=1$, $T=40$, background plateau $n_0=0.65$ and onset $a=15$.
The episode ends at $b=a+L$, where $L$ denotes its duration measured in memory times $1/\beta$.
We simulate 1000 independent catalogues for each combination
\begin{equation}
 n_+\in\{1,1.05,1.10,1.20,1.35,1.50\}, ~
 L\in\{2,4,6,8\}.
\end{equation}
Without the episode, the stationary process with $n_0=0.65$ produces on average $\mu T/(1-n_0)\simeq171$ events in $[0,T]$.  The episode adds events in proportion to its strength and duration: the median catalogue size grows from $178$ events for $(n_+,L)=(1,2)$ to $241$ for $(1,8)$, $753$ for $(1.35,8)$, and $1671$ for $(1.50,8)$, with broad, right-skewed distributions (5--95\% range $407$--$4501$ for $(1.50,8)$).  The quantity that controls detectability is not $N$ but the offspring exposure $E_+$ of the episode [Eq.~\eqref{eq:exposure}], whose median grows from $9$ for $(1,2)$ to $297$ for $(1.35,8)$ and $622$ for $(1.50,8)$ (Table~\ref{tab:sizes}).  The phase diagram thus probes catalogues of a few hundred to a few thousand events, comparable at its lower end to the MERS line list of Sec.~\ref{sec:mers}.
Here the true interval is supplied in order to measure the intrinsic statistical power for estimating $n_+$, separately from the multiple-search problem.  Each catalogue is analysed twice, with $\beta=1$ known and with $\beta$ profiled.

\begin{table}[t]
\caption{\label{tab:sizes}Median number of events $N$ in $[0,T]$ and median offspring exposure $E_+$ of the episode [Eq.~\eqref{eq:exposure}] for the 1000 catalogues of each cell of Fig.~\ref{fig:mc}; entries are $N/E_+$.}
\begin{ruledtabular}
\setlength{\tabcolsep}{3pt}
\begin{tabular}{lcccccc}
 & \multicolumn{6}{c}{$n_+$}\\
$L$ & 1 & 1.05 & 1.10 & 1.20 & 1.35 & 1.50\\
\hline
2 & 178/9 & 184/10 & 184/10 & 187/11 & 200/13 & 208/15\\
4 & 194/24 & 201/27 & 209/30 & 228/36 & 264/46 & 323/65\\
6 & 217/46 & 233/53 & 247/60 & 288/77 & 422/127 & 654/202\\
8 & 241/72 & 262/84 & 297/100 & 400/150 & 753/297 & 1671/622\\
\end{tabular}
\end{ruledtabular}
\end{table}

\begin{figure*}[t]
\includegraphics[width=0.98\textwidth]{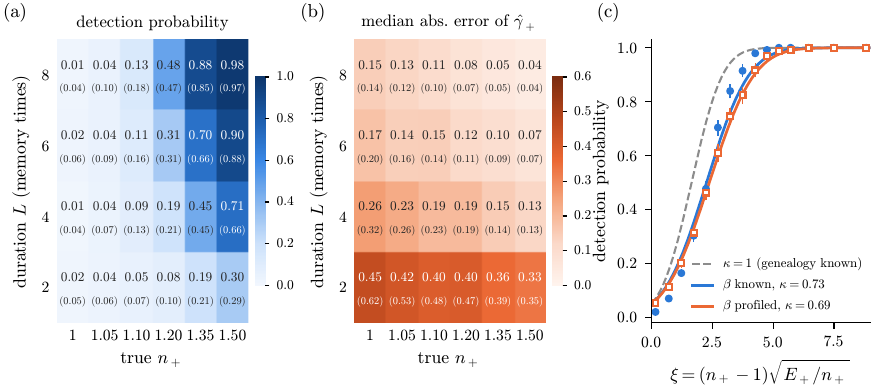}
\caption{\label{fig:mc}Finite-sample detectability of a supercritical plateau, from 1000 catalogues per cell.  Parameters are $\mu=1.5$, $\beta=1$, $T=40$, $n_0=0.65$, and onset $a=15$; the episode $[a,a+L)$ has branching ratio $n_+$, and its window is known.  (a) Empirical probability of rejecting $n_+\le1$ at the one-sided 5\% level.  Each cell gives two values for the same catalogues: the upper number, which also sets the color, is obtained with $\beta=1$ known and the Wald test \eqref{eq:wald}; the lower number, in parentheses, is obtained with $\beta$ estimated by profile likelihood \eqref{eq:profile} and the likelihood-ratio test \eqref{eq:lr} at the threshold $r>1.91$, which gives a 5\% level at $n_+=1$.  (b) Median absolute error of the growth exponent $\hat\gamma_+=\hat\beta(\hat n_+-1)$, with the same layout: $\beta$ known (upper number; equal to the error of $\hat n_+$ since $\beta=1$) and $\beta$ profiled (in parentheses).  (c) Scaling collapse: detection frequency of all 24 cells as a function of the variable $\xi=(n_+-1)\sqrt{E_+/n_+}$ of Eq.~\eqref{eq:xi}, where $E_+=\sum_{t_i\in[a,b)}F_i$ is the offspring exposure of the episode [Eqs.~\eqref{eq:Fi} and~\eqref{eq:exposure}], computed for each catalogue; points are bin averages with 95\% binomial error bars, for $\beta$ known (filled circles) and profiled (open squares).  The dashed curve is the complete-data prediction $\Phi(\xi-1.645)$ ($\Phi$ denotes the standard normal cumulative distribution function), which would hold if the genealogy were observed; the solid curves are one-parameter probit fits $\Phi(\kappa\xi-1.645)$, with $\kappa=0.73$ ($\beta$ known) and $\kappa=0.69$ ($\beta$ profiled).}
\end{figure*}

Figure~\ref{fig:mc}(a) shows the empirical probability that Eq.~\eqref{eq:wald} rejects $n_+\le1$ at the nominal one-sided 5\% level.  Power rises sharply in both $n_+-1$ and $L$.  For $n_+=1.35$, detection increases from $0.19$ at $L=2$ to $0.45$, $0.70$, and $0.88$ at $L=4,6,8$.  For $n_+=1.50$ the corresponding values are $0.30$, $0.71$, $0.90$, and $0.98$.  At $n_+=1$ the empirical rejection frequencies are $1.1$--$2.0\%$, indicating that the simple observed-information Wald diagnostic is conservative in these simulations rather than spuriously declaring supercriticality.

When $\beta$ is profiled, the likelihood-ratio statistic \eqref{eq:lr} is overdispersed under the null in catalogues of this size: at $n_+=1$, its standard deviation is $1.37$, and the nominal test rejects in $6.5$--$7.6\%$ of the catalogues.  The numbers in parentheses in Fig.~\ref{fig:mc}(a) therefore use the threshold $r>1.91$, which gives a 5\% level at $n_+=1$; in applications, the same calibration is provided by a parametric bootstrap from the constrained fit $\widehat\theta_0$.  With this calibration, profiling $\beta$ costs little power: $0.855$ against $0.883$ for $(n_+,L)=(1.35,8)$, and $0.45$ in both cases for $(1.35,4)$.

The estimation error, which we measure on the growth exponent $\widehat\gamma_+=\widehat\beta(\widehat n_+-1)$ because this is the quantity identified when $\beta$ is unknown, decreases with both the excess $n_+-1$ and the duration $L$ [Fig.~\ref{fig:mc}(b)]; for $\beta=1$ known, it coincides with the error on $\widehat n_+$.  At $n_+=1.35$ and $\beta$ known, the median absolute error decreases from $0.36$ at $L=2$ to $0.054$ at $L=8$, where the median estimate is $\widehat n_+=1.325$.  With $\beta$ profiled, the error on $\widehat\gamma_+$ is almost the same for $L\ge6$ ($0.052$ at $L=8$), whereas $\widehat n_+$ is displaced along the ridge: the median $\widehat\beta$ is $0.73$ and the median $\widehat n_+$ is $1.47$, with a median absolute error of $0.20$, four times that obtained with $\beta$ known.  Thus the relevant sample-size variable is not just the total number of events, which itself increases strongly in a supercritical burst, but also the number of memory times over which positive growth persists.

The two control parameters in fact enter through a single combination.  If the genealogy were observed, the numbers of observed children of the parents in the episode would be Poisson with means $n_+F_i$, the complete-data Fisher information on $n_+$ would be $E_+/n_+$, with $E_+=\sum_{t_i\in[a,b)}F_i$ the offspring exposure of the episode, and the one-sided Wald statistic would be
\begin{equation}
 \xi=(n_+-1)\sqrt{E_+/n_+}.
 \label{eq:xi}
\end{equation}
The latent genealogy and the nuisance parameters $(\mu,n_1,n_3)$ remove part of this information, so that $Z\simeq\kappa\,\xi$ with $\kappa<1$, $\kappa^2$ being the fraction of complete-data information retained \cite{DempsterLairdRubin1977}.  Figure~\ref{fig:mc}(c) shows that the detection frequencies of all 24 cells, binned by the value of $\xi$ computed from the exposure of each catalogue, collapse onto the single probit curve $\Pr(Z>1.645)\simeq\Phi(\kappa\xi-1.645)$ with $\kappa=0.73$: roughly half of the complete-data information ($\kappa^2\simeq0.53$) survives the latent attribution.  With $\beta$ profiled and the calibrated threshold, the same collapse holds with $\kappa=0.69$: estimating the memory time removes little of the information on the sign of $n_+-1$.  Cells with very different $(n_+,L)$ but similar $\xi$ have the same power; for instance $(1.20,8)$ and $(1.35,4)$ give $0.48$ and $0.45$, while $(1.35,6)$ and $(1.50,4)$ give $0.70$ and $0.71$.  Equation~\eqref{eq:xi} also quantifies the cost of weak supercriticality: a $50\%$ detection probability requires $E_+\simeq(1.645/\kappa)^2\,n_+/(n_+-1)^2$ exposed parents, about $2\times10^3$ for $n_+=1.05$ and $6\times10^3$ for $n_+=1.03$, whatever the duration of the episode.

\begin{figure}[t]
\includegraphics[width=\columnwidth]{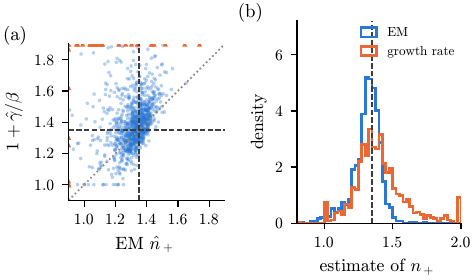}
\caption{\label{fig:growth}Genealogical versus growth-rate estimates of $n_+$ for 1000 catalogues with $n_+=1.35$ and $L=8$ (other parameters as in Fig.~\ref{fig:mc}).  (a) Realization-by-realization scatter; dashed lines mark the true value and the dotted line is the diagonal; the 39 estimates outside the plotted range are drawn as triangles on its boundary.  (b) Distributions of the two estimators.  Both estimators use the known window and $\beta=1$; the growth fit uses Eq.~\eqref{eq:growthfit} with $D$ unconstrained, on bins of width $1/(2\beta)$.  The genealogical estimator is about twice as precise and only weakly correlated with the growth estimate (rank correlation $0.37$).}
\end{figure}

For the moderate case $n_+=1.35$, $L=8$, the genealogical and growth-rate estimators can also be compared across the 1000 realizations (Fig.~\ref{fig:growth}).  The nonlinear fit of Eq.~\eqref{eq:growthfit}, on bins of half a memory time with $D$ unconstrained in sign, converges in all catalogues.  The median genealogical estimate is $1.325$ (interquartile range $1.263$--$1.375$), whereas the median growth estimate is $1.383$ ($1.293$--$1.487$); the root-mean-square errors are $0.14$ and $0.29$, respectively, and the rank correlation between the two estimates is only $0.37$.  Constraining $D\ge0$, as if it were a background offset, raises the median growth estimate to $1.45$, which illustrates the bias discussed after Eq.~\eqref{eq:growthexact}.  We interpret the weak correlation not as a contradiction but as a useful division of labor: genealogy uses the full conditional-event structure and is the principal estimator, whereas transient exponential growth is a lower-efficiency, model-transparent cross-check.

The constant-$n$ fits of Sec.~\ref{sec:recover} are not accidental.  Fitting a stationary model ($n$ constant, $\beta=1$ known) to 300 catalogues of each cell of Fig.~\ref{fig:mc} gives a median $\widehat n=0.96$ (10--90\% range $0.88$--$0.98$) for $(n_+,L)=(1.35,8)$ and $0.98$ ($0.94$--$0.99$) for $(1.50,8)$, whereas the time average of $n(t)$ over the window is $0.79$ and $0.82$; the hypothesis $n\ge1$ cannot be rejected at the 5\% level in $86\%$ and $96\%$ of these catalogues, and the true fraction of triggered events is $0.92$ and $0.97$ (medians).  Averaged by a stationary model, a transient supercritical episode lasting several memory times thus masquerades as a stationary process close to criticality, without any constraint $n<1$ being imposed.

\section{Endogenous versus exogenous episodes}
\label{sec:endoexo}

A burst of events can be produced either by a transient increase of the branching ratio or by a transient increase of the exogenous immigration rate.  The total intensity alone cannot separate the two, since a given profile $\lambda(t)$ can be reproduced by compensating changes of $\mu(t)$ and $n(t)$; an unmodelled non-stationary background biases the branching ratio upward, possibly to apparent criticality \cite{FilimonovSornette2015,Wheatley2019,Wehrli2021}.  What separates the two mechanisms is the lag structure: immigrants are placed independently of the past, whereas offspring follow their parents with the kernel $h_\beta$.  The question is how to let the data exploit this information without prejudging the answer.

\subsection{Competing episodes selected by likelihood}

With a piecewise-constant background $\mu_m$ on intervals $B_m$ of length $|B_m|$, the immigrant probability \eqref{eq:p0} becomes $p_{0j}=\mu_{m(j)}/\lambda_j$, and the background part of the M step \eqref{eq:EM} generalizes to
\begin{equation}
 \mu_m^{\rm new}=\frac{1}{|B_m|}\sum_{t_j\in B_m}p_{0j},
 \label{eq:mubins}
\end{equation}
while the update of $n_r$ is unchanged.  The flexibility of the background is then selected by the information criterion, as in the EM calibration of financial Hawkes processes with a flexible deterministic background \cite{Wheatley2019}; related approaches with time-dependent backgrounds include Refs.~\cite{ChenHall2013,Omi2017}.  
For the transient-episode question we compare four models by the BIC of Eq.~\eqref{eq:bic}, where $k$ is the number of free parameters, each change point counting as one: (i) no episode, with constant $\mu$ and $n$ ($k=2$); (ii) an endogenous episode, with $n$ taking three plateau values separated at the scanned change points $a$ and $b$, and constant $\mu$ ($k=6$); (iii) an exogenous episode, with $\mu$ taking three values separated at scanned change points and constant $n$ ($k=6$); and (iv) both, with common change points ($k=8$).
Model (iii) is the exogenous mirror of model (ii): it gives the exogenous explanation the same number of parameters and the same time resolution as the endogenous one.

Supercriticality is declared only if the selected model contains an endogenous episode with $Z>1.645$. 

\subsection{Kernel-smoothed stochastic reconstruction}

An alternative is to let both functions vary freely and to estimate them by kernel smoothing, in the spirit of stochastic declustering \cite{ZhuangOgataVJ2002,ZhuangOgataVJ2004,ZhuangMateu2019}.  With the immigrant probabilities $p_{0j}$ and the expected numbers of observed children $q_i=\sum_{j>i}p_{ij}$, one iterates
\begin{equation}
 \widehat\mu(t)=\sum_j p_{0j}\,K_{h_\mu}(t-t_j),
~
 \widehat n(t)=\frac{\sum_i K_{h_A}(t-t_i)\,q_i}{\sum_i K_{h_A}(t-t_i)\,F_i},
 \label{eq:kernelrec}
\end{equation}
with smoothing kernels $K_h$ of bandwidths $h_\mu$ and $h_A$, rescaling the amplitudes at each step so that $\int_0^T\widehat\mu\,\dd t=\sum_jp_{0j}$ and $\sum_i\widehat n(t_i)F_i=\sum_j(1-p_{0j})$ \cite{ZhuangMateu2019}.  The second relation in Eq.~\eqref{eq:kernelrec} is the update \eqref{eq:EM} with the regime indicator $\one\{t_i\in I_r\}$ replaced by $K_{h_A}(t-t_i)$: both maximize the expected complete-data log-likelihood of offspring production, $\sum_i w_i\,[q_i\log c-c\,F_i]$, with indicator or kernel weights $w_i$.  Taking $\E[q_i]\simeq n(t_i)F_i$,
\begin{equation}
 \widehat n(t)\simeq\frac{\int K_{h_A}(t-s)\,n(s)\,w(s)\,\dd s}{\int K_{h_A}(t-s)\,w(s)\,\dd s},
 \label{eq:smear}
\end{equation}
where $w(s)=\varrho(s)\,G(T-s)$, $\varrho(s)$ is the event density, and $G$ the cumulative kernel.  The estimator returns a weighted convolution of the true profile with the smoothing kernel \cite{FanGijbels1996}.  Inside a supercritical episode, $w(s)$ grows like $e^{\gamma(s-a)}$; near the onset, the estimate is therefore pulled towards the subcritical level by the pre-episode events; after the termination, it remains pulled towards $n_+$ by the dense late-episode events, and the plateau is attenuated unless $h_A\ll L$.  The change points, which are the objects of interest here, are not parameters of Eq.~\eqref{eq:kernelrec}; they can only be read off a smoothed curve, with $O(h_A)$ resolution and no error bar.

More fundamentally, the kernel formulation adds a time scale to a problem that already has two: the memory $1/\beta$, which carries the information separating offspring from immigrants, and the episode duration $L$.  The bandwidth $h_\mu$ fixes \emph{a priori} how fast the exogenous rate may vary.  If $h_\mu\lesssim L$, the background can follow the burst and absorbs endogenous events; if $h_\mu\gg L$, a genuine exogenous surge cannot be followed and is pushed into $\widehat n(t)$.  In the change-point formulation, the corresponding scales, the durations of the endogenous and exogenous episodes, are parameters selected by the likelihood together with all others.

\subsection{Monte Carlo comparison}
\label{sec:endoexomc}

We compare the estimators on three scenarios with $\beta=1$, $T=40$, and 200 catalogues each: (B) the endogenous episode of Sec.~\ref{sec:detect}, with $\mu=1.5$, $n_0=0.65$, and $n_+=1.35$ on $[15,23)$; (B$_0$) the null $n\equiv0.65$; and (C) an exogenous surge, with $n\equiv0.65$ and $\mu(t)=1.5+6\cdot\one\{15\le t<23\}$, i.e.\ a five-fold increase of immigration on the same window.  The change-point EM is used either with the endogenous model only (Sec.~\ref{sec:detect}) or with the four-model selection above.  In this section, $\beta$ is treated as known for all methods, including the kernel reconstruction, in order to isolate the competition between background and fertility; the rates of the endogenous-only scan with $\beta$ profiled or misspecified are given in Sec.~\ref{sec:robust}.  The kernel reconstruction uses a Gaussian kernel on a grid of step $0.2$, a flat initialization, and 380 iterations, with $h_\mu=h_A=h\in\{1,2,3,4,6\}$ and with wide backgrounds $h_\mu\in\{10,20\}$ combined with $h_A\in\{1,2,4\}$.  Since Eq.~\eqref{eq:kernelrec} carries no test, we report the naive rule $\max_t\widehat n(t)>1$; its threshold could be recalibrated by simulation, but only under an assumed alternative, which is what is in general unknown in applications.

\begin{table}[t]
\caption{\label{tab:endoexo}Fraction of catalogues in which a supercritical episode is declared, for an endogenous episode (B), the null (B$_0$), and an exogenous surge (C), 200 catalogues each; the last column (ampl.) is the median of $\hat n_+$ (EM) or of the window average of $\hat n(t)$ (kernel) in (B), where $n_+=1.35$.  EM declares an episode when $\Delta\mathrm{BIC}<0$ and $Z>1.645$; the kernel rule is $\max_t\hat n(t)>1$, with bandwidths $h=h_\mu=h_A$ or $(h_\mu,h_A)$.}
\begin{ruledtabular}
\begin{tabular}{lcccc}
 & (B) & (B$_0$) & (C) & ampl.\\
\hline
EM, endogenous only & 0.78 & 0.00 & 0.025 & 1.30\\
EM, endo/exo choice & 0.71 & 0.00 & 0.005 & ---\\
Kernel, $h=1$ & 0.16 & 0.17 & 0.17 & 0.33\\
Kernel, $h=2$ & 0.49 & 0.13 & 0.095 & 0.90\\
Kernel, $h=3$ & 0.63 & 0.09 & 0.08 & 1.01\\
Kernel, $h=4$ & 0.67 & 0.06 & 0.04 & 1.01\\
Kernel, $h=6$ & 0.61 & 0.005 & 0.01 & 0.99\\
Kernel, $(10,1)$ & 0.97 & 0.53 & 0.67 & 1.16\\
Kernel, $(10,2)$ & 0.94 & 0.17 & 0.34 & 1.14\\
Kernel, $(10,4)$ & 0.81 & 0.035 & 0.08 & 1.06\\
Kernel, $(20,2)$ & 0.93 & 0.095 & 0.37 & 1.15\\
\end{tabular}
\end{ruledtabular}
\end{table}

\begin{figure*}[t]
\includegraphics[width=0.98\textwidth]{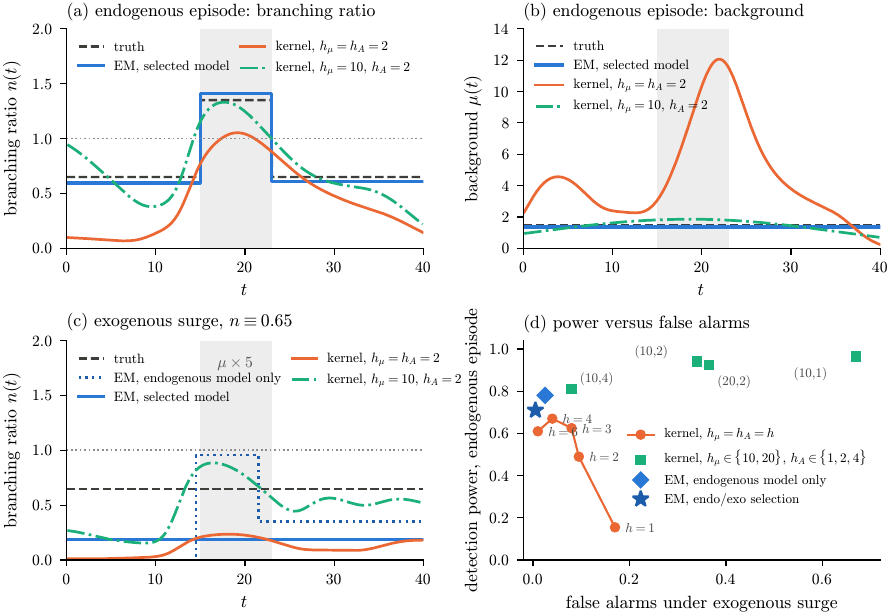}
\caption{\label{fig:endoexo}Endogenous versus exogenous episodes ($\beta=1$, $T=40$).  (a,b) Catalogue of scenario (B) ($N=777$; $n_+=1.35$ on the shaded window, $\mu=1.5$): branching ratio and background estimated by the change-point EM (selected model: endogenous episode on $[15,23)$ with $\hat n_+=1.41\pm0.07$) and by the kernel reconstruction with two bandwidth choices.  With $h_\mu=h_A=2$, the background absorbs the burst.  (c) Catalogue of scenario (C) closest to the median size ($N=308$): $n\equiv0.65$ with a five-fold exogenous surge on the shaded window.  The EM restricted to the endogenous model finds a transient plateau $\hat n_+=0.96$, below unity; the four-model selection prefers the exogenous episode, whose constant $\hat n=0.19$ is depressed because part of the aftershock tail of the surge is assigned to the background.  (d) Detection power in (B) versus false alarms in (C), 200 catalogues each, for the kernel reconstruction over nine bandwidth settings (naive rule $\max_t\hat n(t)>1$) and for the two EM procedures ($\Delta\mathrm{BIC}<0$ and $Z>1.645$).}
\end{figure*}

Table~\ref{tab:endoexo} and Fig.~\ref{fig:endoexo} summarize the results.  The change-point EM restricted to the endogenous model detects the episode of (B) in $78\%$ of the catalogues and never declares supercriticality under the null; under the exogenous surge, it does so in $2.5\%$ of the catalogues, so the lag structure already protects it substantially.  Adding the exogenous alternative reduces these false alarms to $0.5\%$ at the price of some power ($71\%$): $12.5\%$ of the endogenous bursts are then attributed to an exogenous episode, the honest cost of admitting the competing explanation.  We also note that, in such short catalogues, the EM estimate of a constant background is biased upward (median $\widehat\mu=1.97$ for $\mu=1.5$ under the null), with a corresponding downward bias of the subcritical plateaus; this small-sample ridge between immigration and low-level triggering does not affect the supercritical plateau.

The kernel reconstruction behaves very differently [Fig.~\ref{fig:endoexo}(a,b)].  With $h_\mu=h_A=2$, the background follows the burst: in (B) its median value inside the window is $9.1$, six times the true rate, while the median window average of $\widehat n(t)$ is $0.90$, so that $\widehat n(t)$ exceeds unity in only $49\%$ of the catalogues.  Widening the background bandwidth restores power, to $94\%$ for $h_\mu=10$, $h_A=2$, but the same setting declares a supercritical episode in $34\%$ of the catalogues of (C) and in $17\%$ under the null.  Across the nine bandwidth settings, the kernel results trace a trade-off curve between power and false alarms [Fig.~\ref{fig:endoexo}(d)], and both change-point procedures lie above and to the left of it.  The kernel setting that matches the false-alarm rate of the EM has lower power ($61\%$ for $h=6$, with $1\%$ false alarms under the surge), and the settings that exceed the power of the EM do so with false-alarm rates three to twenty-seven times larger ($8\%$ for $h_\mu=10$, $h_A=4$, and $67\%$ for $h_\mu=10$, $h_A=1$).  The amplitude is attenuated as well: the median window average of $\widehat n(t)$ never exceeds $1.16$ for any bandwidth, against $\widehat n_+=1.30$ for the EM and $n_+=1.35$.

These results support the time-scale argument made above.  The change-point formulation lets the likelihood decide where and for how long the endogenous or exogenous episode acts; the kernel formulation fixes the time resolution of both functions a priori, and no single choice is adequate for both alternatives.

\subsection{Unknown memory time, misspecification, and near-critical nulls}
\label{sec:robust}

Table~\ref{tab:robust} reports the blind scan of Sec.~\ref{sec:recover}, with the endogenous model only and the decision rule $\Delta\mathrm{BIC}<0$ and $r>1.645$, when $\beta$ is profiled, fixed at its true value, or misspecified by a factor of two in either direction.  Besides the episode (B) and the surge (C) of Sec.~\ref{sec:endoexomc}, it includes stationary nulls with $n\equiv0.90$ and $0.95$ and the critical case $n\equiv1$.  These catalogues are three to seventeen times larger than under $n\equiv0.65$ (median $N=543$, $1187$, and $2818$, against $164$) and are dominated by long endogenous cascades.  Such a cascade produces a burst of activity that resembles a supercritical episode, although the branching ratio never exceeds unity; these nulls therefore test whether the scan mistakes the large fluctuations of a near-critical stationary process for a transient crossing of $n=1$.

\begin{table}[t]
\caption{\label{tab:robust}Blind scan with the memory time profiled (prof.), known ($\beta=1$, the true value in all simulations), or misspecified (0.5 and 2), on a scan grid of step 4 refined to step 1.  Detection requires $\Delta\mathrm{BIC}<0$ and $r>1.645$, with the endogenous model only.  Episode (B): detection rate, and medians of $\hat n_+$ (true $1.35$) and of $\hat\gamma_+=\hat\beta(\hat n_+-1)$ (true $0.35$); 200 catalogues with $\beta$ profiled, 100 otherwise.  False-alarm rates for stationary nulls $n\equiv n_0$ (100 catalogues for $n_0=0.65$, 50 otherwise) and for the exogenous surge (C), 100 catalogues.}
\begin{ruledtabular}
\setlength{\tabcolsep}{2.5pt}
\begin{tabular}{lcccccccc}
 & \multicolumn{3}{c}{episode (B)} & \multicolumn{4}{c}{null, $n_0=$} & (C)\\
$\beta$ & det. & $\hat n_+$ & $\hat\gamma_+$ & 0.65 & 0.90 & 0.95 & 1.00 & \\
\hline
prof. & 0.65 & 1.39 & 0.30 & 0 & 0 & 0 & 0 & 0.04\\
1 & 0.79 & 1.33 & 0.33 & 0 & 0 & 0 & 0 & 0.04\\
0.5 & 0.83 & 1.67 & 0.33 & 0.02 & 0 & 0 & 0.02 & 0.34\\
2 & 0.46 & 1.12 & 0.24 & 0 & 0 & 0 & 0 & 0\\
\end{tabular}
\end{ruledtabular}
\end{table}

None of the 150 near-critical and critical null catalogues (50 for each of $n\equiv0.90$, $0.95$, and $1.00$) produces a detection, whether $\beta$ is profiled or fixed at its true value.  Zero detections in 150 catalogues bound the false-alarm rate of each of the two procedures below $2\%$ at 95\% confidence, since $1-0.05^{1/150}\simeq0.020$.  In the critical case, the transient model is preferred by BIC in $30\%$ of the catalogues, which reflects the genuine nonstationarity of a critical process, whose mean rate grows linearly in time, but the selected plateau is then never significantly above unity.  The burstiness of a near-critical stationary process is therefore not mistaken for a transient crossing.

A misspecified memory time, by contrast, is not innocuous, and the direction of the error matters.  Too long a memory ($\beta=0.5$) attributes the immigrants of an exogenous surge to earlier events: false alarms under (C) rise from $4\%$ to $34\%$, and the higher power in (B), $83\%$, is produced by the same mechanism.  Too short a memory ($\beta=2$) attributes late offspring to immigration; it is conservative but nearly halves the power, to $46\%$.  Profiling $\beta$ removes both distortions: under the surge, the median $\widehat\beta$ is $1.11$ and the false-alarm rate is that of the correctly specified model.

The price of estimating $\beta$ is a loss of power, from $79\%$ to $65\%$, and of precision on $n_+$ but much less on $\gamma_+$.  The loss of power arises in the model-selection step rather than in the test: the transient model is preferred by BIC in $70\%$ of the catalogues against $83\%$ with $\beta$ known, because a constant-$n$ model with a free memory time absorbs part of the burst with a short memory (median $\widehat\beta=2.16$), whereas with a known window the calibrated test loses almost no power (Sec.~\ref{sec:detect}).  Across the 200 catalogues of (B), the median $\widehat\beta$ is $0.79$ (interquartile range $0.44$--$1.21$), and $\widehat n_+$ spreads along the ridge, with median $1.39$ ($1.15$--$1.77$) against $1.33$ ($1.22$--$1.38$) when $\beta$ is known, whereas the median growth exponent is $\widehat\gamma_+=0.30$ ($0.15$--$0.37$) against $0.33$ ($0.22$--$0.38$), for a true value of $0.35$.  The fixed-$\beta$ rows display the same invariance: halving $\beta$ doubles $\widehat n_+-1$ but leaves $\widehat\gamma_+$ unchanged.  What a short transient record measures is thus the growth exponent of the episode, whose sign is that of $n_+-1$; the branching ratio itself requires the memory time, which must come from external information, such as the serial interval in Sec.~\ref{sec:mers}, or from substantially longer records.


\section{Empirical application: transient supercriticality in the 2015 Korean MERS outbreak}
\label{sec:mers}

We test the reconstruction on a real epidemic for which the onset times and transmission history are unusually well documented.  We use the public line list of the 2015 Middle East respiratory syndrome coronavirus (MERS-CoV) outbreak in the Republic of Korea distributed in the \texttt{outbreaks} data archive \cite{OutbreaksData} (\url{https://www.reconverse.org/outbreaks/reference/mers_korea_2015.html}).  The archive contains 162 cases in this snapshot; 135 have a recorded symptom-onset date between 11 May and 15 June 2015 and are used here.  The outbreak was initiated by a single imported index case and was subsequently amplified almost entirely by nosocomial transmission \cite{KCDC2015MERS}.  We therefore condition on the earliest-onset case as the known immigrant and apply the genealogical EM reconstruction to the remaining 134 cases, rather than estimating a continuing exogenous immigration rate.  This is the one-cluster specialization of Eq.~\eqref{eq:EM}: every later case is assigned probabilistically to a preceding parent, while the regime-specific offspring exposures retain the exact finite-window correction.

The triggering-time density is taken from the independently estimated serial-interval distribution.  Cowling \emph{et al.} estimated a mean serial interval of 12.6 days and a standard deviation of 2.8 days for the Korean outbreak \cite{Cowling2015MERS}.  We therefore use the normalized Gamma kernel
\begin{align}
 h(u)&=\frac{u^{k-1}e^{-u/s}}{\Gamma(k)s^k}\one_{u>0},
 \label{eq:merskernel}\\
 k&=\left(\frac{12.6}{2.8}\right)^2=20.25,
 ~
 s=\frac{2.8^2}{12.6}=0.622~\mathrm{d}.
\end{align}
Fixing the time scale of the kernel from independent data removes the ridge of Sec.~\ref{sec:beta}, so that the magnitudes of the cohort fertilities, and not only their position relative to unity, are meaningful conditional on the serial interval; we examine this condition below with an alternative serial interval.  Because onset times are reported only to the day, tied events are independently jittered uniformly within their reporting day.  We first scan a single fertility change point $b$, the minimal model of a crossing,
\begin{equation}
 n(t)=
 \begin{cases}
 n_{\rm early},&t<b,\\
 n_{\rm late},&t\ge b,
 \end{cases}
 \label{eq:mers2regime}
\end{equation}
and compare it with a constant-$n$ model by BIC; we then allow a second change point.  Each complete change-point scan is repeated over 100 (one change) or 20 (two changes) independent within-day jitterings, and the plateau estimates at the selected boundaries are summarized over a further 300 to 500 jitterings.

\begin{figure*}[t]
\includegraphics[width=0.98\textwidth]{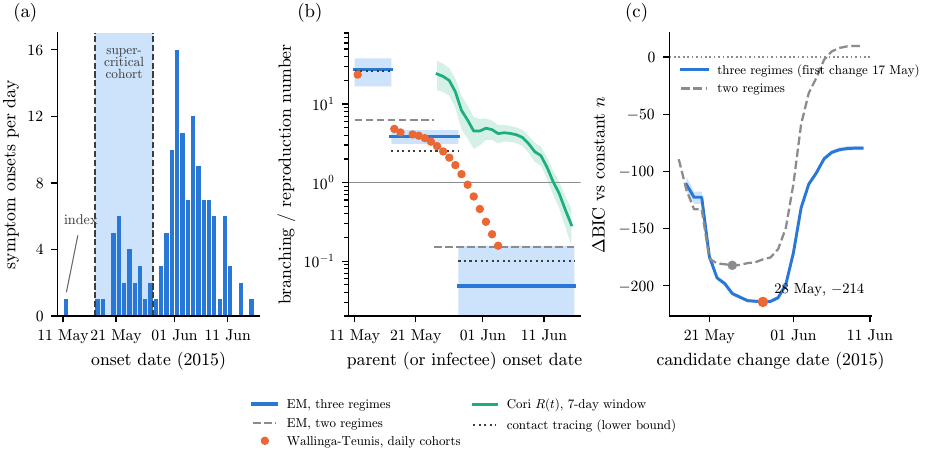}
\caption{\label{fig:mers}
Genealogical reconstruction of the 2015 MERS-CoV outbreak in the Republic of Korea, using the 135 cases with symptom onset between 11 May and 15 June 2015, a Gamma serial-interval kernel with mean 12.6 days and standard deviation 2.8 days \cite{Cowling2015MERS}, and the first onset as the imported immigrant.  (a) Daily symptom-onset incidence; the dashed lines mark the two change points selected by BIC (17 and 28 May), and the shaded region the supercritical parent cohort between them.  (b) Branching ratio of parent cohorts on a logarithmic scale: three-regime EM estimate with conditional 95\% Wald bands (solid), one-change estimate of Eq.~\eqref{eq:mersfit} (dashed), case reproduction numbers of daily onset cohorts from the Wallinga--Teunis attribution with right-truncation correction (dots; cohorts with at least half of their offspring observable), lower bounds from the documented contact-tracing pairs (dotted), and the instantaneous reproduction number of Cori \emph{et al.} on 7-day windows, indexed by the infectee onset date and shown from 24 May (line with 95\% credible band).  The horizontal line marks $n=1$.  (c) Median $\Delta\mathrm{BIC}$ relative to the constant-$n$ model over within-day jitterings, as a function of the candidate change date: one change point (dashed; minimum $-182$ on 24 May) and a second change point with the first fixed at 17 May (solid with interquartile band; minimum $-214$ on 28 May).}
\end{figure*}

The results are shown in Fig.~\ref{fig:mers}.  A constant branching ratio gives $\widehat n=1.77$, which would suggest that the outbreak remained globally supercritical throughout the observation window.  Allowing one change point produces a qualitatively different conclusion.  The BIC minimum occurs for parent onset on 24 May, with
\begin{equation}
 \widehat n_{\rm early}=6.28,
~~
 \widehat n_{\rm late}=0.153,
~~
 \Delta\mathrm{BIC}=-182,
 \label{eq:mersfit}
\end{equation}
relative to the constant-$n$ model.  Conditional observed-information standard errors are 0.58 and 0.072, giving approximate 95\% intervals $[5.15,7.41]$ and $[0.011,0.294]$, respectively.  The conclusion is insensitive to the unresolved ordering of cases within each day: the 100 complete jittered scans have median change date 24 May, with 95\% of selected dates between 22 and 25 May, while at the fixed 24 May boundary, the 5--95\% jitter ranges are $6.23$--$6.31$ for $n_{\rm early}$ and $0.139$--$0.169$ for $n_{\rm late}$.

A second change point improves the fit further, by $\Delta\mathrm{BIC}=-32$ relative to the one-change model in all 20 jittered scans ($\Delta\mathrm{BIC}\simeq-214$ relative to constant $n$).  The selected boundaries are 17 and 28 May (17 of 20 scans; 17 and 27 May in the other three).  They define three parent cohorts, with symptom onset before 17 May, from 17 to 27 May, and from 28 May onward, whose fertilities we denote by $n_{\rm index}$, $n_{17\text{--}27\,{\rm May}}$, and $n_{\ge28\,{\rm May}}$; the first cohort contains only the imported index case (onset on 11 May), hence its label.  The estimates are
\begin{equation}
 \begin{aligned}
 \widehat n_{\rm index}&=27.5\pm5.5,\\
 \widehat n_{17\text{--}27\,{\rm May}}&=3.87\pm0.39,\\
 \widehat n_{\ge28\,{\rm May}}&=0.048\pm0.055,
 \end{aligned}
 \label{eq:mers3fit}
\end{equation}
with conditional observed-information standard errors and 5--95\% jitter ranges $26.8$--$28.1$, $3.83$--$3.90$, and $0.038$--$0.061$.  The estimated fertility of the index case agrees with the 26 secondary cases attributed to him by contact tracing \cite{KCDC2015MERS}. The one-change model \eqref{eq:mersfit} thus averaged this single superspreader with a genuinely supercritical cohort of hospital transmission, $\widehat n\simeq3.9$, whose fertility collapsed for parents with symptom onset from 28 May.  This is the date on which the mean delay from symptom onset to laboratory confirmation fell from 8.1 to 4.4 days \cite{Cowling2015MERS}, after the index case had been confirmed on 20 May and the investigation had been widened to further hospitals \cite{KCDC2015MERS}.

\paragraph*{Validation against contact tracing.}  The \texttt{outbreaks} archive also lists 98 documented transmission pairs, covering $73\%$ of the 134 secondary cases in the window.  Dividing the documented children of each cohort by its exposure gives the lower bounds $26$, $2.49$, and $0.10$ for the three regimes, consistent with Eq.~\eqref{eq:mers3fit} given that about a quarter of the transmissions are undocumented.  For the children of documented parents in the first two regimes, the EM assigns on average $87\%$ and $94\%$ of the attribution probability to the correct cohort.  For the five documented children of late parents, it assigns only $17\%$, whereas the non-iterated Wallinga--Teunis attribution assigns $75\%$: once a strongly supercritical cohort has been identified, the fertility-weighted attribution transfers ambiguous offspring of the following cohort to it.  The estimate $\widehat n_{\ge28\,{\rm May}}$ is therefore biased low, although both it and the documented bound are far below unity.  The documented pairs also give a serial interval of $14.5\pm5.2$ days, longer and broader than the $12.6\pm2.8$ days of Ref.~\cite{Cowling2015MERS}.  With this kernel, the constant-$n$ fit gives $2.09$, the single change point moves to 21 May ($\Delta\mathrm{BIC}=-195$), and the two change points to 17 and 25 May, with cohort fertilities $33.4$, $5.0$, and $\simeq0$.  The crossing of the parent cohorts through $n=1$ is therefore located in the last week of May, between 21 and 28 May depending on the serial interval.

\paragraph*{Heterogeneity.}  The magnitude of these fertilities should not be interpreted as a homogeneous biological reproduction number.  The Korean MERS outbreak was dominated by extreme transmission heterogeneity: $83.2\%$ of documented transmission events were linked to five superspreaders \cite{KCDC2015MERS}, three of whom, with 38, 21, and 4 documented secondary cases, belong to the supercritical cohort of 17--27 May.  In the present reduced model, $n(t)$ is therefore an effective cohort-averaged Hawkes fertility.  Because the Wald intervals assume Poisson offspring numbers, we checked the supercriticality of this cohort against superspreading \cite{LloydSmith2005,KucharskiAlthaus2015}: if its 27 parents had a mean fertility $n\le1$ with negative-binomial offspring numbers of dispersion $k$, the probability that they produce the 104 children attributed to them would be $1.0\times10^{-5}$ for $k=0.25$ and $1.7\times10^{-3}$ even for the extreme value $k=0.1$.  The supercritical character of this cohort is therefore not an artefact of the Poisson assumption.

\paragraph*{Comparison with standard reproduction-number estimators.}  Figure~\ref{fig:mers}(b) overlays the case reproduction numbers of daily onset cohorts computed with the non-iterated Wallinga--Teunis attribution \cite{WallingaTeunis2004} and the right-truncation correction of Ref.~\cite{Cauchemez2006}, together with the instantaneous reproduction number estimated with the method of Cori \emph{et al.} \cite{Cori2013} on sliding 7-day windows, with the same serial-interval distribution and a Gamma prior of mean and standard deviation 5.  The Wallinga--Teunis case numbers decline gradually, from $3.3$--$4.8$ for the cohorts of 17--23 May to below unity from 29 May; aggregated over the two cohorts of Eq.~\eqref{eq:mersfit}, they give $4.9$ and $0.65$ instead of $6.28$ and $0.153$, which reflects the sharpening of the attribution by the EM.  All cohort-indexed estimates thus date the collapse of transmission potential to the last week of May.  The indicators that are usually monitored tell a very different story.  Incidence keeps rising until 1 June, and the instantaneous reproduction number, indexed by the onset of the infectee, remains significantly above unity until 11 June, its posterior mean falling below unity only on 13 June, about two weeks after the collapse.  This offset is not an error of these estimators: indexing by the infectee instead of the infector shifts the signal by one serial interval, to which half the smoothing window adds \cite{Fraser2007}.  Its consequences are nevertheless practical.  Read as a measure of when the outbreak came under control, the instantaneous reproduction number would credit measures taken in early June with a change that the parent-cohort reconstruction places at 28 May, when the delay from symptom onset to laboratory confirmation was halved.  The incidence peak of 1 June is likewise not a signature of subcriticality: it is the delayed echo, one serial interval later, of the supercritical cohort of 17--27 May.

\paragraph*{Retrospective dating versus real-time detection.}  A cohort-indexed estimate requires the offspring of the cohort to be observed, so the crossing can be dated precisely only in hindsight.  Repeating the blind two-change scan on data truncated at successive dates, with the exposures $F_i$ recomputed for each truncation, a supercritical regime following the index case is established with data up to 1 June, the day of the incidence peak, and the subcriticality of the most recent parent cohorts with data up to 11 June (upper 95\% bound $0.89$), two to three days before the 7-day instantaneous reproduction number crosses below unity.  This is about one serial interval after the crossing, which is close to the earliest possible detection by any method, since the fertility of a cohort is revealed only by its offspring. The genealogical reconstruction therefore does not anticipate the incidence peak; its value lies in dating and testing the change of transmission potential by parent cohort, with a modest lead over smoothed instantaneous estimates, whereas a constant-$n$ calibration would have reported sustained supercriticality throughout.  The coincidence of the inferred crossing with the acceleration of diagnosis on 28 May is consistent with an effect of the interventions, which we do not identify causally.


\section{Discussion}\label{sec:discussion}

\subsection{Stationarity, non-explosion, and apparent criticality}
\label{sec:apparent}

The conditions
\begin{equation}
 n<1,\qquad N(T)<\infty\ \text{a.s.},\qquad \gamma>0
\end{equation}
answer three different questions: the first is a condition for stationarity, the second for finite-time non-explosion, and the third characterizes a supercritical growth mode.  For the standard exponential Hawkes process, $n>1$ violates stationarity but not finite-horizon existence.  If the reproduction rate later falls below one, because of intervention, resource depletion, behavioral adaptation, stress release, or any other mechanism, a finite supercritical excursion is a natural model object rather than a pathology.  Classical and locally stationary theory rightly imposes subcriticality when the goal is an equilibrium process, finite stationary moments, or a local stationary approximation \cite{BremeaudMassoulie1996,Roueff2016,MammenMueller2023}, whereas supercritical theory studies exponential growth and scaling beyond the stability boundary \cite{HelmstetterSornette2002,LiuXuZhang2026}.  The statistical gap lies between them: a real record is finite, regimes change, and the scientific question may be whether the system temporarily crossed the boundary.  The estimator must therefore allow the crossing before the data can decide whether it occurred.

Excluding the crossing is not a neutral choice.  A stationary model fitted to a record that contains a transient supercritical episode returns a branching ratio just below unity, and the fraction of triggered events, often used as a proxy for $n$, is bounded by one by construction (Sec.~\ref{sec:detect}).  Averaged by a stationary model, transient supercriticality is read as near-criticality.  This mechanism adds to the apparent criticality produced by an unmodelled nonstationary background \cite{FilimonovSornette2015,Wheatley2019}, and the two can be told apart only by a model that admits both an endogenous and an exogenous episode, as in Sec.~\ref{sec:endoexo}.  It also echoes the sweeping of an instability as an alternative to self-organized criticality \cite{Sornette1994}: in both cases, the signatures of criticality come from a system that passes through or near the critical point rather than one that resides at it.  The near-critical branching ratios reported for financial markets \cite{FilimonovSornette2012,Hardiman2013}, and for other systems claimed to operate at criticality \cite{Munoz2018}, are therefore candidates for a time-resolved reanalysis.  Our simulations show that the mechanism is possible, and the test of Sec.~\ref{sec:beta} makes it falsifiable.  A constant-$n$ fit can also err in the opposite direction: in the MERS outbreak, it reported sustained supercriticality, $\widehat n=1.77$, although the parent cohorts had been subcritical since 28 May (Sec.~\ref{sec:mers}).  In both cases, a stationary calibration averages the regimes with weights set by the event density, so that its result reflects the history of the record rather than its current state.

\subsection{Genealogy, change points, and smoothing}
\label{sec:genealogy}

The genealogical formulation used here is well suited to this problem because it makes the allocation between endogenous and exogenous events explicit.  The total intensity alone can be reproduced by compensating changes in background and excitation, an ambiguity documented in financial Hawkes calibration \cite{Wheatley2019,Wehrli2021}.  Posterior parent probabilities do not add observed Fisher information, but they provide the natural complete-data representation for estimating offspring production, and therefore organize the inference around the quantity whose crossing matters.  Our results also explain why aggressive smoothing can be counterproductive.  Kernel smoothing of the branching ratio attenuates a real fertility jump and smears its onset and termination, and a smoothing bandwidth imposed on the background either absorbs the jump into the exogenous rate or creates a spurious one (Sec.~\ref{sec:endoexo}).  Change-point or total-variation formulations, with the relevant time scales selected by the likelihood, are preferable when abrupt regime changes are part of the alternative hypothesis.

\subsection{Conditions for identifiability and limits of detection}
\label{sec:identifiability}

Beyond reconstruction, the question is when transient supercriticality is identifiable at all from a single finite record.  Our results delineate three conditions.

\paragraph*{(i) Information.}  Detection requires the episode to carry enough exposed parents: the test statistic scales as $\kappa\xi$ with $\xi=(n_+-1)\sqrt{E_+/n_+}$ [Eq.~\eqref{eq:xi}], so that a $50\%$ detection probability requires $E_+\simeq(1.645/\kappa)^2\,n_+/(n_+-1)^2$, whatever the duration of the episode, with $\kappa\simeq0.7$ whether $\beta$ is known or estimated.  Weak or brief excursions are therefore undetectable in typical catalogues, and the absence of a detection is not evidence of subcriticality.

\paragraph*{(ii) Separation of scales.}  Fertility and immigration are distinguished only through the lag structure of the triggering kernel.  An exogenous surge lasting several memory times is separated from an endogenous episode (Sec.~\ref{sec:endoexo}), but background variations on the memory scale $1/\beta$, or backgrounds of unrestricted flexibility, can reproduce a burst without any change of $n$.  In that regime, the allocation is not identified by the data and must be settled by external information.

\paragraph*{(iii) Model dependence.}  In a short transient record, the data identify the growth exponent $\gamma_+=\beta(n_+-1)$ of the episode rather than $n_+$ itself, whose magnitude is conditional on the kernel.  A misspecified memory time is not harmless: too long a memory turns an exogenous surge into a spurious supercritical episode in a third of the catalogues (Table~\ref{tab:robust}).  Including $\beta$ in the calibration removes this bias and preserves the test, because the sign of $\gamma_+$ is that of $n_+-1$, at the cost of some power in the selection of the window; $\widehat n_+$ then inherits the large uncertainty on $\widehat\beta$, and a reliable magnitude requires external information on the kernel (Sec.~\ref{sec:robust}).  The kernel \emph{shape} is not tested here: an exponential kernel fitted to long-memory, Omori-type triggering, common in seismicity, may redistribute offspring across regimes, and in the MERS application the choice of serial-interval distribution shifts the dated crossing by up to a week (Sec.~\ref{sec:mers}).  For long-memory kernels, the growth-rate cross-check also loses most of its power, because the growth exponent vanishes as $(n-1)^{1/\theta}$ near criticality (Appendix~\ref{app:omori}).

Two further limitations apply.  Our alternatives are piecewise constant; gradual crossings can be handled by local basis functions or fused or total-variation penalties, at the price of a resolution-dependent definition of the onset.  And the BIC scan combined with the likelihood-ratio test is a practical selection device rather than a complete theory of inference after selection.  For a known window and $\beta$ profiled, the likelihood-ratio statistic is overdispersed in short catalogues ($7\%$ rejections at the nominal $5\%$ level, Sec.~\ref{sec:detect}), so that its threshold should be calibrated by a parametric bootstrap.  Combined with the BIC scan, the procedure had false-alarm rates at or below the nominal level for subcritical, near-critical, and critical nulls and for an exogenous surge, with $\beta$ profiled or known (Table~\ref{tab:robust}); it is therefore conservative in the cases studied, which does not guarantee that it is so in general.  A systematic map of identifiability under misspecification (kernel shape, background flexibility on the memory scale, unknown windows) is a statistical problem in its own right, which we leave for future work.

\subsection{A protocol for empirical applications}
\label{sec:protocol}

These conditions suggest a concrete protocol.  
\begin{enumerate}
\item First, fit a genealogical model that permits $n(t)$ to cross unity, with the kernel parameters taken from independent data when available and otherwise included in the calibration.  
\item Second, compare a subcritical-only fit against a transient-supercritical alternative \emph{and} against an exogenous-episode alternative of equal flexibility.  
\item Third, test $n_+>1$ with a likelihood ratio calibrated by parametric bootstrap, and check that the episode carries enough exposure for a detection to be possible at all.  
\item Fourth, propagate the uncertainty in parentage, change points, and kernel.  
\item Fifth, test whether the inferred supercritical interval carries the independent growth signature predicted by the fitted memory kernel. 
\end{enumerate}
 Only after these steps should an observed burst be interpreted as endogenous supercritical amplification rather than as exogenous forcing or model misspecification.

In the MERS application of Sec.~\ref{sec:mers}, the kernel is taken from independent data (step one), and the exogenous alternative of step two is settled by external information, since the outbreak was seeded by a single documented importation; the supercritical cohort is detected far above the threshold of step three ($\xi\simeq7.6$); and step five is not informative there, because the supercritical cohort spans less than one serial interval, too short for a Malthusian growth mode to develop.

\section{Conclusion}\label{sec:conclusion}

We have stressed that a branching ratio above one need not be excluded from a finite Hawkes record.  For the exponential kernel, it produces a finite, exponentially amplifying process on every bounded horizon; the prohibition $n<1$ belongs to stationary theory.  Once this distinction is made, transient supercriticality becomes an estimable dynamical regime.  
We have shown that genealogical EM reconstruction identifies the offspring production associated with different parent-time regimes, finite-window exposure corrects the end of the catalogue, and a likelihood scan can locate the interval without prespecifying it.  The simulation phase diagram shows where this is statistically possible: weak or very short crossings are hard to distinguish from criticality, whereas sustained excursions are recovered with high probability, and detectability is governed by the single variable $\xi=(n_+-1)\sqrt{E_+/n_+}$.  When the memory time is unknown, a short record identifies the growth exponent $\beta(n_+-1)$, hence the sign of $n_+-1$, but not the magnitude of $n_+$.  Conversely, a stationary fit to a record containing a transient supercritical episode returns a branching ratio just below one, so that transient supercriticality can masquerade as near-criticality; allowing $n(t)$ to cross unity is therefore also a way of testing claims of criticality. Letting the likelihood choose between an endogenous and an exogenous episode protects the endogenous interpretation against surges of immigration; kernel smoothing with a bandwidth fixed a priori offers no such protection.  Our proposed framework therefore replaces an a priori stability constraint by a testable hypothesis about finite-time dynamics.

\begin{acknowledgments}
The simulation code used for the numerical results in this manuscript is available upon reasonable request.  We thank Jiancang~Zhuang for discussions and for a critical reading of an earlier version that prompted the robustness and identifiability analyses of Secs.~\ref{sec:robust} and~\ref{sec:identifiability}, and Weiqiang Wang for his preliminary investigations.
\end{acknowledgments}

\appendix
\section{Growth rate for the Omori kernel}
\label{app:omori}

The growth estimator \eqref{eq:growth} is exact for an exponential kernel in a constant supercritical regime; for a general kernel it must be replaced by the Malthusian equation \eqref{eq:malthus}.  For the normalized Omori kernel
\begin{equation}
 g(t)=\frac{\theta c^\theta}{(t+c)^{1+\theta}},
 \label{eq:omori}
\end{equation}
the exact relation is
\begin{equation}
 \frac{1}{n}=1-e^{\gamma c}(\gamma c)^\theta\,\Gamma(1-\theta,\gamma c).
 \label{eq:omorimalthus}
\end{equation}
For the ETAS-relevant range $0<\theta<1$ and close to criticality,
\begin{equation}
 \gamma\simeq\frac{1}{c}\left[\frac{n-1}{n\,\Gamma(1-\theta)}\right]^{1/\theta},
 \label{eq:omorigamma}
\end{equation}
so that the growth rate vanishes nonlinearly as $n\downarrow1$, $\gamma\propto(n-1)^{1/\theta}$.  For the commonly encountered value $\theta=0.2$ and a mildly supercritical branching ratio $n=1.1$, Eq.~\eqref{eq:omorigamma} gives $\gamma c\simeq[0.1/(1.1\,\Gamma(0.8))]^5\simeq2.9\times10^{-6}$, in agreement with the exact solution of Eq.~\eqref{eq:omorimalthus}, $\gamma c\simeq2.90\times10^{-6}$.  Even though the process is already $10\%$ above the critical branching ratio, its asymptotic exponential growth rate is extraordinarily small because of the long Omori memory.  The $\gamma$-based estimator is therefore least informative precisely in the empirically important near-critical, long-memory ETAS regime, which strengthens the motivation for the genealogical method; exogenous background growth can, moreover, mimic endogenous supercritical growth unless it is modeled separately.

\end{document}